\documentclass[pra,twocolumn,tightenlines,showpacs,nofootinbib,floatfix]{revtex4}
\usepackage{multirow}
\usepackage{amsmath}
\usepackage{bm}
\usepackage{dcolumn}
\usepackage{graphicx}

\newcommand{\eref}[1]{Eq.~(\ref{#1})}
\newcommand{\tref}[1]{Table~\ref{#1}}

\begin{document}
\title{Relativistic many-body calculations of van der Waals coefficients for Yb-Li and Yb-Rb dimers}

\author{S.~G.~Porsev$^{1,2}$}
\author{M.~S.~Safronova$^{1,3}$}
\author{A. Derevianko$^4$}
\author{Charles W. Clark$^3$}
\affiliation{ $^1$Department of Physics and Astronomy, University of Delaware,
    Newark, Delaware 19716, USA\\
$^2$Petersburg Nuclear Physics Institute, Gatchina, Leningrad District, 188300, Russia \\
$^3$Joint Quantum Institute, National Institute of Standards and Technology and the \\
University of Maryland, Gaithersburg, Maryland, 20899-8410, USA\\
$^4$Physics Department, University of Nevada, Reno, Nevada 89557, USA }
\date{\today}

\begin{abstract}
We derive the relativistic formulas for the van der Waals coefficients of Yb--alkali dimers that correlate to
ground and excited separated-atom limits. We calculate   $C_6$ and $C_8$
coefficients of particular experimental interest. We also derive a semi-empirical formula that expresses the $C_8$ coefficient of heteronuclear
$A+B$ dimers in terms of the $C_6$ and $C_8$ coefficients of  homonuclear dimers and the static dipole and quadrupole
polarizabilities of the atomic states $A$ and $B$. We report
results of calculation of the $C_6$ coefficients for the Yb--Rb  $^3\!P_1^o+5s\, ^2\!S_{1/2}$ and $^1\!S_0+5p\, ^2\!P^o_{1/2}$ dimers,
and the $C_8$ coefficients for the Yb--Li $^1\!S_0+2s\, ^2\!S_{1/2}$ and Yb--Rb $^1\!S_0+5s\, ^2\!S_{1/2}$ dimers.
Uncertainties are estimated for all predicted properties.
\end{abstract}
\maketitle

\section{Introduction}

The subject of long-range interactions of Yb atoms and Yb-alkali atoms  has recently became of much interest owing to
study of quantum gas mixtures
\cite{KatSugShi13l,BruHut12,BruHut13,TakSaiTak12l,KatYamShi12l,IvaKhrHan11l,BorCiuJul11l,HarTakYam11l,HanKhrDow11l,
OkaHarMur10l,TakKomHon04l,TojKitEno06l,BorCiuJul09l,EnoKitToj08,KitEnoKas08l,YamTaiSug13l,NemBauMun09l,MunBruMad11,BauMunGor11l},
development of optical lattice clocks~\cite{JiaLudLem11l,LemSteShe11l}, study of fundamental symmetries \cite{MeyBoh09},
quantum computing \cite{ShiKatYam09l} and practical realization of quantum simulation
proposals \cite{MicBreZol06,ReiJulDeu09,SanOdoJav11}. Yb is a particularly suitable candidate for all of these applications
owing to its five bosonic and two fermionic stable isotopes in natural abundance, the $^1\!S_0$ ground state, the long-lived
metastable $6s6p\,\, ^3\!P_0^o$ state, and transitions at convenient wavelengths for laser cooling and trapping.

Presently,  there is much interest in forming cold molecules with both electric and magnetic dipole
moments because of the greater possibilities for trapping and manipulation ~\cite{IvaKhrHan11l,HarTakYos11,NemBauMun09l,MunBruMad11,BruHut12}.
Unlike
ultracold alkali dimer molecules~\cite{NiOspMir08l} with largely diamagnetic ground states, the alkali-Yb dimers possess
unpaired electron spin, and thereby can be controlled by both electric and magnetic fields. This added control enlarges the
class of many-body Hamiltonians that can be  simulated with ultracold molecules~\cite{MicBreZol06}. The  characterization of
Yb interactions with alkali atoms is crucial  for selecting efficient pathways for assembling Yb-alkali molecules via
photo- or magneto-association
techniques~\cite{BruHut12,BruHut13}. Predictions of magnetically tunable Feshbach
resonance positions and widths were recently reported for Yb--Li, Yb--Rb, and Yb--Cs \cite{BruHut12,BruHut13}.

Quantum degenerate mixtures of Li and Yb were realized in \cite{HanKhrDow11l,HarTakYam11l}
using sympathetic cooling of Li atoms by evaporatively cooled Yb atoms.
Controlled production of ultracold YbRb* molecules by photoassociation in a mixture of Rb and Yb gases was
recently reported in \cite{NemBauMun09l,MunBruMad11}.
In particular, Ref. \cite{NemBauMun09l} explored production of ultracold
Yb--Rb $^1\!S_0-5p\,\, ^2\!P^o_{1/2}$ dimers by photoassociation in a mixture of Rb and Yb gases.
The spectroscopic investigation of vibrational levels in the
electronic ground state of the $^{176}$Yb$^{87}$Rb molecule  was recently carried out in~\cite{MunBruMad11}.
Unusually strong interactions in a thermal mixture of $^{87}$Rb and $^{174}$Yb ultracold atoms which
caused a significant modification of the spatial distribution were observed in~\cite{BauMunGor11l}.

This brings urgency to understanding the collisional interactions of Yb, both among its various isotopes and with other atoms,
in particularly Li
\cite{BruHut12,IvaKhrHan11l,HarTakYam11l,HanKhrDow11l,OkaHarMur10l} and Rb \cite{BruHut13,MunBruMad11,BauMunGor11l,NemBauMun09l}.
Knowledge of the $C_6$ and $C_8$ long-range interaction coefficients in Yb--alkali
dimers is critical to understanding the physics of dilute gas mixtures for the applications mentioned above.
Recently, we evaluated the $C_6$ and $C_8$ coefficients for the Yb--Yb ($^1\!S_0 +\,^1\!S_0$) dimer and
found them to be $C_6=1929(39)$~\cite{SafPorCla12} and  $C_8=1.88(6) \times 10^5$~\cite{PorSafDer13l}, in excellent
agreement with the experimental results $C_6=1932(35)$ and  $C_8=1.9(5) \times 10^5$~\cite{KitEnoKas08l}.
However, the expressions for the $C_6$ and $C_8$ coefficients used there cannot be applied to
calculations for the Yb--Rb ($^3\!P_1^o+ 5s\,\, ^2\!S_{1/2}$) and
($^1\!S_0+5p\,\, ^2\!P^o_{1/2}$) dimers due to
the presence of the Yb $^3\!P_1^o-\,^1\!S_0$ and Rb $5p\,\, ^2\!P^o_{1/2} - 5s\,\, ^2\!S_{1/2}$ decay channels and different angular couplings.

In this work we derive relativistic expressions for the $C_6$ coefficients of heteronuclear dimers
involving excited state atoms with strong decay channels to the ground state.
The non-relativistic formalism has been described in Refs.~\cite{MarDal95,MarSad99,ZhaMit07}.
We apply the resulting formulas to evaluate
the $C_6$ coefficients for the ($^3\!P_1^o+5s\,\, ^2\!S_{1/2}$) and ($^1\!S_0+5p\,\, ^2\!P^o_{1/2}$) dimers. We also
evaluate the $C_8$ coefficients for the Yb--Li ($^1\!S_0+2s\,\, ^2\!S_{1/2}$) and  Yb--Rb ($^1\!S_0+5s\,\, ^2\!S_{1/2}$) dimers.

For the case when $A$ and $B$ are the spherically symmetric atomic states
and there are no downward transitions from either of these states,
we derive a semi-empirical formula for the $C_8$ coefficient of
heteronuclear ($A+B$) dimers following a method suggested by Tang~\cite{Tan69}.
The resulting expression allows us to evaluate this property using the $C_6$ and $C_8$ coefficients of both homonuclear
($A+A$) and ($B+B$) dimers and static dipole $\alpha_1(0)$ and quadrupole $\alpha_2(0)$ polarizabilities of the atoms $A$ and $B$.
We find that this semi-empirical formula gives the same value of $C_8$ for Yb--Li and Yb--Rb as our \textit{ab initio}
numerical calculation within our estimated uncertainties.

We have evaluated the  uncertainties of all quantities calculated in this work.
The results obtained here can be used for the analysis of existing measurements and for
planning future experiments. The expressions for the $C_6$ and $C_8$ coefficients derived in this work
as well as the methodology of calculations can be used for evaluation of van der Waals coefficients in
similar systems.

The paper is organized as follows. In Sections~\ref{GenForm} and~\ref{approx} we present the general formalism
and derive the analytical expressions for the $C_6$ coefficients for the
($^3\!P_1^o+ 5s\,\, ^2\!S_{1/2}$) and ($^1\!S_0+5p\,\, ^2\!P^o_{1/2}$) dimers. In Section~\ref{method} we briefly
describe the method of calculations. Finally, Section~\ref{results} is devoted to discussion of the results
and contains concluding remarks.

 Unless stated otherwise, we use atomic units (a.u.) for all matrix
elements and polarizabilities throughout this paper: the numerical values of the elementary
 charge, $|e|$, the reduced Planck constant, $\hbar = h/2
\pi$, and the electron mass, $m_e$, are set equal to 1. The atomic unit for polarizability can be
converted to SI units via $\alpha/h$~[Hz/(V/m)$^2$]=2.48832$\times10^{-8}\alpha$~(a.u.), where the
conversion coefficient is $4\pi \epsilon_0 a^3_0/h$ and the Planck constant $h$ is factored out in
order to provide direct conversion into frequency units; $a_0$ is the Bohr radius and $\epsilon_0$
is the electric constant.

\section{General formalism}
\label{GenForm}
We investigate the molecular potentials that asymptotically correlate to separated
$|A\rangle$ and $|B\rangle$ atomic states. The general formalism for homonuclear dimers has been discussed in ~\cite{PorSafDer13l} and
we only give a brief outline here.
We take the Rb atom to be in the state $|J_A M_A \rangle$ and Yb atom to be
in the state $|J_B M_B \rangle$. The model space for treatment of Yb-Rb dimer consists of the product states
\begin{equation}
\Psi_\Omega = |J_A M_A \rangle \, |J_B M_B \rangle ,
\label{WF_YbRb}
\end{equation}%
where $\Omega =M_A + M_B$ is the sum of the projections of the total angular momenta $M_A$ and $M_B$ on the internuclear axis.
We assume that $\Omega$ is a good quantum number for all Yb--Rb dimers studied here, i.e. the coupling scheme can be described
by the Hund's case (c). The correct molecular wave functions can be obtained by diagonalizing the molecular
Hamiltonian
\begin{equation}
\hat{H}=\hat{H}_A + \hat{H}_B+\hat{V}(R)
\label{Eq_Hamilt}
\end{equation}%
in the model space, where $\hat{H}_A$ and $\hat{H}_B$ represent the Hamiltonians of the two noninteracting
atoms.  $\hat{V}(R)$ is the residual electrostatic potential defined as the
full Coulomb interaction energy of the dimer excluding interactions of the
atomic electrons with their parent nuclei.

The multipole expansion of the potential $V(R)$ is given by
\begin{equation}
V(R)=\sum_{l,L=0}^{\infty }V_{lL}/R^{l+L+1}\, ,
\label{e1}
\end{equation}
where $V_{lL}$ are given in general form in~\cite{DalDav66}.
We restrict our consideration by the dipole-dipole and
dipole-quadrupole interactions in the second-order perturbation theory.
The first two terms of the expansion given by Eq.~(\ref{e1}) are
\begin{eqnarray}
V_{dd}(R) =-\frac{1}{R^3} \sum_{\mu=-1}^1 w_{\mu}^{(1)} (d_{\mu})_A (d_{-\mu })_B,
\label{Vdd}
\end{eqnarray}
\begin{eqnarray}
V_{dq}(R) = \frac{1}{R^4} && \sum_{\mu=-1}^1 w_{\mu }^{(2)}
\left[ (d_{\mu})_A (Q_{-\mu })_B \right. \nonumber \\
&& \left. - (Q_{\mu })_A (d_{-\mu })_B \right] ,
\label{Vdq}
\end{eqnarray}
where $d$ and $Q$ are the dipole and quadrupole operators, respectively. The dipole and quadrupole weights are
\begin{eqnarray}
w_{\mu }^{(1)} &\equiv& 1+\delta _{\mu 0} , \nonumber \\
w_{\mu }^{(2)} &\equiv& \frac{6}{\sqrt{\left( 1-\mu \right) !\left( 1+\mu
\right) !\left( 2-\mu \right) !\left( 2+\mu \right) !}}.
\label{wmu}
\end{eqnarray}%
Numerically $w_{-1}^{(2)}=$ $w_{+1}^{(2)}=\sqrt{3}$ and $w_{0}^{(2)}=3$.

The resulting dispersion potential can be approximated as
\begin{equation}
U(R) \approx -\frac{C_6(\Omega)}{R^6}-\frac{C_8(\Omega)}{R^8} ,
\label{UR1}
\end{equation}%
where the second-order corrections, associated with the $C_6$ and $C_8$ coefficients,
are given by
\begin{eqnarray}
\frac{C_{6}(\Omega)}{R^6} &=&
\sum_{\Psi_i \neq \Psi_{\Omega}}
\frac{ \langle \Psi_{\Omega} |\hat{V}_{dd}| \Psi_{i} \rangle
       \langle \Psi_i |\hat{V}_{dd}| \Psi_{\Omega} \rangle} {E_i -\mathcal{E}} ,
\label{C6R6}
\end{eqnarray}
\begin{eqnarray}
\frac{C_8(\Omega)}{R^8} &=& \sum_{\Psi_{i} \neq \Psi_{\Omega}}
\frac{\langle \Psi_{\Omega}|\hat{V}_{dq}| \Psi_i \rangle
      \langle \Psi_i |\hat{V}_{dq}| \Psi_{\Omega} \rangle}{E_i-\mathcal{E}} .
\label{C8R8}
\end{eqnarray}
and the intermediate molecular state $|\Psi_{i}\rangle $ with unperturbed
energy $E_{i}$ runs over a \emph{complete} set of two-atom states, excluding
the model-space states, Eq.~(\ref{WF_YbRb}). The energy $\mathcal{E}$ is given by
 $\mathcal{E} \equiv E_A+E_B$, where $E_A$ and $E_B$ are
the atomic energies of the $|A\rangle$ and $|B\rangle$ states.
 The complete set of doubled atomic states meets the condition
$
\sum_{\Psi_i} |\Psi _{i}\rangle \langle \Psi _{i}| = 1.
$

Using Eqs.~(\ref{WF_YbRb}), (\ref{Vdd}), and (\ref{C6R6}),
we separate the angular and radial parts of the $C_6$ coefficient and,
after some transformations, arrive at the expression
\begin{equation}
C_6(\Omega)=\sum_{j=|J_A-1|}^{J_A+1}\, \sum_{J=|J_B-1|}^{J_B+1} A_{j J}(\Omega)\, X_{j J} ,
\label{C6_YbRb}
\end{equation}
where
\begin{eqnarray}
A_{j J}(\Omega) &=& \sum_{\mu m M} \left\{ w^{(1)}_{\mu} \left(
\begin{array}{ccc}
 J_A &  1  & j \\
-M_A & \mu & m
\end{array}
\right) \right. \nonumber \\
&\times& \left. \left(
\begin{array}{ccc}
 J_B & 1 & J \\
-M_B & -\mu & M%
\end{array}
\right) \right\}^2 ,
\label{AjJ}
\end{eqnarray}
with $\Omega= M_A + M_B = m + M$.

The quantities $X_{j J}$ are given by
\begin{eqnarray}
&&X_{j J} = \label{XjJ} \\
&& \sum_{\gamma_n, \gamma_k} \frac{|\langle A ||d|| \gamma_n, J_n=j \rangle|^2\,
|\langle B ||d|| \gamma_k, J_k=J \rangle|^2} {E_n - E_A + E_k - E_B} ,
\nonumber
\end{eqnarray}
where the total angular momenta of the intermediate atomic states, $J_n$ and $J_k$, are assumed to be
fixed and equal to $j$ and $J$, respectively; $\gamma_n$ and $\gamma_k$ include all
the quantum numbers except the total angular momenta $J_n$ and $J_k$.

If $A$ and $B$ are spherically symmetric atomic states and there are no downward transitions
from either of them, we can apply the Casimir-Polder identity,
\begin{eqnarray}
\frac{1}{x+y} = \frac{2}{\pi} \int_0^\infty d\omega \frac{x}{x^2+\omega^2}
\frac{y}{y^2+\omega^2}; \,\, x > 0,\, y>0 ,
\label{CP}
\end{eqnarray}
to simplify the general expressions. For the (A+B) dimer, we obtain the well known $C_6$ and $C_8$ coefficient formulas
(see, e.g.,~\cite{PatTan97})
\begin{eqnarray}
C^{AB}_6 &=& C^{AB}(1,1), \nonumber \\
C^{AB}_8 &=& C^{AB}(1,2)+C^{AB}(2,1) ,
\label{vdW}
\end{eqnarray}
where the coefficients $C^{AB}(l,L)$ ($l,L=1,2$) are
quadratures of electric-dipole, $\alpha_1(i \omega)$, and electric-quadrupole,
$\alpha_2(i \omega)$, dynamic polarizabilities at an imaginary frequency:
\begin{eqnarray}
C^{AB}(1,1) = \frac{3}{\pi}\, \int_0^\infty\, \alpha_1^A(i \omega)\,
\alpha_1^B(i \omega)\, d\omega,
\label{CAB_11}
\end{eqnarray}
\begin{eqnarray}
C^{AB}(1,2) &=& \frac{15}{2\pi}\, \int_0^\infty\, \alpha_1^A(i \omega)\,
\alpha_2^B(i \omega)\, d\omega , \nonumber \\
C^{AB}(2,1) &=& \frac{15}{2\pi}\, \int_0^\infty\, \alpha_2^A(i \omega)\,
\alpha_1^B(i \omega)\, d\omega .
\label{C_12}
\end{eqnarray}

The derivation of the formulas for the $C_6$ coefficients for the Yb--Rb  $^1\!S_0+5p\, ^2\!P^o_{1/2}$ and $^3\!P_1^o+5s\, ^2\!S_{1/2}$
dimers, which have strong downward transitions, and resulting final expressions 
are given in Sections~\ref{5p1S0} and~\ref{5s3P1}.

\section{Semiempirical expressions for $C_6$ and $C_8$ coefficients}
\label{approx}
Using the method suggested by Tang~\cite{Tan69}, we are able to derive approximate
formulas for the $C^{AB}_6$ and $C^{AB}_8$ coefficients. Following Tang, we
express the dynamic $2^K$-pole polarizability of a state $X$ at the imaginary values of the frequency $i\omega$
as
\begin{eqnarray}
\alpha_K^X (i\omega) &=& \sum_n \frac{f_{nX}}{\omega_{nX}^2 + \omega^2} \nonumber \\
&=& \sum_n \frac{f_{nX}}{\omega^2_{nX}}\, \frac{1}{1 + (\omega/\omega_{nX})^2} ,
\label{alpha_K1}
\end{eqnarray}
where $f_{nX}$ are the oscillator strengths. Tang's procedure involves approximating Eq.~(\ref{alpha_K1}) by
\begin{eqnarray}
\alpha_K^X (i\omega) &\approx& \frac{\alpha_K^X(0)}{1+(\omega\,\phi_K^X)^2} .
\label{alpha_K}
\end{eqnarray}
Here, $\phi_K^X$ is a free parameter that is determined below.

If we substitute Eq.~(\ref{alpha_K}) in (\ref{CAB_11}), we find ~\cite{Tan69}
\begin{eqnarray}
\phi_1^X = \frac{3}{4}\,\frac{(\alpha_1^X(0))^2}{C_6^{XX}},
\label{eta1}
\end{eqnarray}
where $X$ is $A$ or $B$.
Using this expression, we find that the $C_6$ coefficient for a
heteronuclear dimer can be approximated by the following formula~\cite{Tan69}
\begin{eqnarray}
C^{AB}_6 \approx \frac{2\, \alpha_1^A(0)\, \alpha_1^B(0)\, C^{AA}_6\,
C^{BB}_6} {C^{BB}_6 (\alpha_1^A(0))^2 + C^{AA}_6 (\alpha_1^B(0))^2 } ,
\label{C6approx}
\end{eqnarray}
where $\alpha_1^A(0)$ and $\alpha_1^B(0)$ are the electric dipole static
polarizabilities of the atomic states $A$ and $B$, and $C^{AA}_6$ and
$C^{BB}_6$ are the $C_6$ coefficients for the ($A+A$) and ($B+B$) dimers, respectively.

To derive the corresponding approximate formula for the $C_8$ coefficient, we use
Eqs.~(\ref{C_12}), (\ref{alpha_K}), and (\ref{CP}).
We obtain
\begin{eqnarray}
C_8^{AB} \approx \frac{15}{4} \left[ \frac{\alpha_1^A(0)\,\alpha_2^B(0)}{\phi_1^A + \phi_2^B}
+ \frac{\alpha_2^A(0)\,\alpha_1^B(0)} {\phi_2^A + \phi_1^B} \right] .
\label{C8approx}
\end{eqnarray}
The quantity $\phi_1^X$ is given by Eq.~(\ref{eta1}) and $\phi_2^X$
can be determined as follows.
For a homonuclear $X+X$ dimer, we have from~Eq.~(\ref{C8approx})
\begin{eqnarray}
C_8^{XX} \approx \frac{15}{2}\, \frac{\alpha_1^X(0)\,\alpha_2^X(0)}{\phi_1^X+ \phi_2^X} .
\label{C8_XX}
\end{eqnarray}
If the $C_8^{XX}$ coefficient is known, then we obtain from Eq.~(\ref{C8_XX})
\begin{eqnarray}
\phi_2^X &\approx& \frac{15}{2}\, \frac{\alpha_1^X(0)\,\alpha_2^X(0)}{%
C_8^{XX}} - \phi_1^X  \nonumber \\
&=& \frac{15}{2}\, \frac{\alpha_1^X(0)\,\alpha_2^X(0)}{C_8^{XX}} - \frac{3}{4%
}\,\frac{(\alpha_1^X(0))^2}{C_6^{XX}} .
\end{eqnarray}
Therefore, the heteronuclear $C_8^{AB}$ coefficient can be obtained using~\eref{C8approx} if one knows the following
quantities: $\alpha_1^A(0)$, $\alpha_2^A(0)$, $\alpha_1^B(0)$, $\alpha_2^B(0)$, $C_6^{AA}$, $C_8^{AA}$, $C_6^{BB}$, and $C_8^{BB}$.

This approximate formula reproduces the result of the first-principle calculations carried out in this work
within the uncertainty estimates, as discussed in Section~\ref{5s1S0}. We suggest that the semiempirical formula can be used
to estimate $C_8$ coefficients for other systems, such as the Yb-Cs dimer.
\section{Method of calculation}
\label{method}
The dynamic electric-dipole and electric-quadrupole Rb polarizabilities, needed to carry out calculations
of the $C_6$ and $C_8$ coefficients, were obtained elsewhere~\cite{DerPorBab10,PorDer03,ZhuDalPor04}.

We calculated the Yb properties needed for this work using the method that combines configuration interaction (CI) and the coupled-cluster
all-order approach (CI+all-order) that treats both core and valence correlation to
all orders~\cite{Koz04,SafKozJoh09,SafKozCla11}. This approach has been demonstrated to give accurate results
for energies, transition properties, and polarizabilities for a variety of divalent and trivalent neutral atoms
and ions ~\cite{SafKozJoh09,SafKozCla11,SafPorCla12,SafMaj13,SafSafPor13}. The application of this method for  Yb has been discussed in
\cite{SafPorCla12,PorSafDer13l} and we give only a brief outline below.

We start from solving the Dirac-Fock (DF) equations,
\[
\hat H_0\, \psi_c = \varepsilon_c \,\psi_c,
\]
where $H_0$ is the relativistic DF
Hamiltonian~\cite{DzuFlaKoz96b,SafKozJoh09} and $\psi_c$ and $\varepsilon_c$ are the
single-electron wave functions and energies; the self-consistent procedure was carried out for
the [$1s^2,...,4f^{14}$] closed core.

The wave functions and the low-lying energy levels are determined by solving
the multiparticle relativistic equation for two valence electrons~\cite%
{KotTup87},
\begin{equation}
H_{\mathrm{eff}}(E_n) \Phi_n = E_n \Phi_n.
\end{equation}
The effective Hamiltonian is defined as
\begin{equation}
H_{\mathrm{eff}}(E) = H_{\mathrm{FC}} + \Sigma(E),
\end{equation}
where $H_{\mathrm{FC}}$ is the Hamiltonian in the frozen-core approximation
and the energy-dependent operator $\Sigma(E)$ accounts for the virtual
core excitations.

The CI space spans $6s-20s$, $6p-20p$, $5d-19d$, $5f-18f$, and $5g-11g$ orbitals
and is effectively complete.
The operator $\Sigma(E)$ is constructed using the linearized coupled-cluster
single-double method~\cite{SafKozJoh09}. For all-order terms evaluation
we use a finite B-spline basis set,
consisting of $N=35$ orbitals for each partial wave with $l\leq5$ and
formed in a spherical cavity with radius 60 a.u.

The dynamic polarizability of the $2^K$-pole operator $T^{(K)}$, $d_{\mu} \equiv T_{\mu}^{(1)}$, $Q_{\mu} \equiv T_{\mu}^{(2)}$,
at imaginary argument is calculated as the sum of the valence and core polarizabilities
\begin{equation}
\alpha_K(i\omega) = \alpha_K^v(i\omega) + \alpha_K^c(i\omega) . 
\end{equation}%
The core polarizability $\alpha_K^c(i\omega)$ includes small $vc$ part that  restores the Pauli principle.
The valence part of the dynamic polarizability, $\alpha_K^{v}(i\omega )$, of
an atomic state $|\Phi \rangle $ is determined by solving the inhomogeneous
equation in the valence space \cite{KozPor99}.
The core contributions to multipole polarizabilities are evaluated in the  relativistic
random-phase approximation (RPA).

The uncertainties of the $C_6$ and $C_8$ coefficients may be expressed via
uncertainties in the static multipole polarizabilities of the atomic states
$A$ and $B$ ($A \neq B$) (see Refs.~\cite{PorDer03,PorSafDer13l} for more detail):
\begin{eqnarray}
 \delta C^{AB}(l,L)  =
 \sqrt{\left( \delta \alpha_l^A(0) \right)^2 +
       \left( \delta \alpha_L^B(0)\right)^2} \, ,
\label{C_AB_l}
\end{eqnarray}
and
\begin{eqnarray}
\Delta C^{AB}_6 &=&  \Delta C^{AB}(1,1) , \nonumber \\
\Delta C^{AB}_8 &=& \sqrt{ (\Delta C^{AB}(1,2))^2+  (\Delta C_{AB}(2,1))^2 } .
\label{C_AB}
\end{eqnarray}
Here, prefixes $\delta$ and $\Delta$  stand for the fractional and absolute
uncertainties, respectively.

We discuss the results of calculations and evaluation of
the uncertainties of the $C_6$ and $C_8$ coefficients in the next section.
For brevity, we use shorter notations for the Li ground state, $2s\,\, ^2\!S_{1/2} \equiv 2s$,
and for the Rb ground and excited states,
$5s\,\, ^2\!S_{1/2} \equiv 5s$ and $5p\,\, ^2\!P^o_{1/2} \equiv 5p_{1/2}$.
\section{Results and discussion}
\label{results}
\subsection{Yb--Li ($6s^2\, ^1\!S_0 + 2s$) and Yb--Rb ($6s^2\, ^1\!S_0 + 5s$) dimers}
\label{5s1S0}
The $C_6$ coefficient for the ground state case $^{1}\!S_{0}+5s$ was previously calculated  in Ref.~\cite{SafPorCla12}.
The $C_8$ coefficient  is calculated in the present work. Since we are considering the dimers with
Yb, Li and Rb in the ground states, the $C_6$ and $C_8$ coefficients are given by Eqs.~(\ref{vdW})-(\ref{C_12}).
\begin{table*}[tbp]
\caption{The values of the static electric-dipole, $\protect\alpha_1$, and
electric-quadrupole, $\protect\alpha_2$, polarizabilities (in a.u.) for the Li, Rb,
and Yb ground states and the $C_6$ and $C_8$ coefficients for
the Li--Li ($2s+2s$), Rb--Rb ($5s+5s$), Yb--Yb ($^1\!S_0 +\, ^1\!S_0$), Yb--Li ($^1\!S_0+2s$) and
Yb--Rb ($^1\!S_0+5s$) dimers. The $C_6$ and $C_8$
coefficients for the Yb--Li ($^1\!S_0+2s$) and  Yb--Rb ($^1\!S_0+5s$) dimers are found by (i) using
the dynamic polarizabilities and Eqs.~(\ref{vdW}--\ref{C_12})
and (ii) using the approximate formulas~(\protect\ref{C6approx}) and
(\protect\ref{C8approx}). The uncertainties are given in parentheses.}
\label{C8aB}%
\begin{ruledtabular}
\begin{tabular}{lllll}
        &                         & (i) Numerical                         & (ii) Approximate    & Other results \\
\hline
Li--Li & $\alpha_1(2s)$           & 164.0(1)\footnotemark[1]              &                     & 164.1125(5)\footnotemark[2] \\
       & $\alpha_2(2s)$           & 1424(4)\footnotemark[3]               &                     & 1423.266(5)\footnotemark[4] \\
       & $C_6(2s+2s)$             & 1389(2)\footnotemark[1]               &                     & 1393.39(16)\footnotemark[4] \\
       & $C_8(2s+2s)$             & $8.34(4)\times 10^4$\footnotemark[3]  &                     &$8.34258(4)\times 10^4$\footnotemark[4]\\[0.3pc]

Rb--Rb & $\alpha_1(5s)$           & 318.6(6)\footnotemark[1]              &                     &  322(4)\footnotemark[5] \\
       & $\alpha_2(5s)$           & 6520(80)\footnotemark[3]              &                     &  6525(37)\footnotemark[5]\\
       & $C_6(5s+5s)$             & 4690(23)\footnotemark[1]              &                     &   \\
       & $C_8(5s+5s)$             & $5.77(8) \times 10^5$\footnotemark[3] &                     &   \\[0.3pc]

Yb--Yb & $\alpha_1(^1\!S_0)$      & 141(2)\footnotemark[6]                &                     &  141(6)\footnotemark[7]\\
       & $\alpha_2(^1\!S_0)$      & 2560(80)\footnotemark[8]              &                     &  \\
       & $C_6(^1\!S_0+\,^1\!S_0)$ & 1929(39)\footnotemark[6]              &                     &  1932(35)\footnotemark[9] \\
       & $C_8(^1\!S_0+\,^1\!S_0)$ & $1.88(6) \times 10^5$\footnotemark[5] &                     & $1.9(5) \times 10^5$\footnotemark[9]\\[0.3pc]

Yb--Li & $C_6(^1\!S_0+2s)$        & 1551(31)\footnotemark[6]              &                     &  1594\footnotemark[10] \\
       & $C_8(^1\!S_0+2s)$        & $1.27(3) \times 10^5$                 & $1.27 \times 10^5$  & \\[0.3pc]

Yb--Rb & $C_6(^1\!S_0+5s)$        & 2837(57)\footnotemark[6]              &  2814               &  2830\footnotemark[10] \\
       &                          &                                       &                     & 2837(13)\footnotemark[11]\\
       & $C_8(^1\!S_0+5s)(1,2)$   & $1.351(24) \times 10^5$               & $1.335 \times 10^5$ &  \\
       & $C_8(^1\!S_0+5s)(2,1)$   & $1.848(57) \times 10^5$               & $1.865 \times 10^5$ &  \\
       & $C_8(^1\!S_0+5s)$        & $3.200(65) \times 10^5$               & $3.199 \times 10^5$ & $4.9(6)\times 10^5$\footnotemark[11]
\end{tabular}
\end{ruledtabular}
\footnotemark[1]{Ref.~\cite{DerPorBab10}};
\footnotemark[2]{Refs.~\cite{PucKcePac11,PucKcePac12e}};
\footnotemark[3]{Ref.~\cite{PorDer03}};
\footnotemark[4]{Ref.~\cite{YanBabDal88}};
\footnotemark[5]{Ref.~\cite{SafSaf11}};
\footnotemark[6]{Ref.~\cite{SafPorCla12}};
\footnotemark[7]{Ref.~\cite{DzuDer10}};
\footnotemark[8]{Ref.~\cite{PorSafDer13l}};
\footnotemark[9]{Ref.~\cite{KitEnoKas08l}};
\footnotemark[10]{Ref.~\cite{BruHut13}, approximate formula};
\footnotemark[11]{Ref.~\cite{BorZucCiu13}}.
\end{table*}

The integrals over $\omega$ needed for the evaluation of $C_6$ and $C_8$ are calculated using
Gaussian quadrature of the integrand computed on a finite grid of discrete imaginary frequencies~\cite{BisPip92,DerPorBab10}.
For example, the integral in the expression for $C^{AB}_6$ coefficient given by Eq.~(\ref{CAB_11}) is replaced by a finite sum
\begin{equation}
C^{AB}_6=\frac{3}{\pi}\sum_{k=1}^{N_g} W_k\, \alpha_1^A{(i\omega_k)}\, \alpha^B_1{(i\omega_k)}
\label{finsum}
\end{equation}
over values of $\alpha_1^{A}(i \omega_k)$ and $\alpha_1^{B}(i \omega_k)$ tabulated at certain frequencies $\omega_k$
yielding an $N_g$-point quadrature, where each term in the sum is weighted by factor $W_k$. In this work we use
points and weights listed in Table A of Ref.~\cite{DerPorBab10} and $N_g=50$.

The Li and Rb ground state dynamic electric-dipole and electric quadrupole polarizabilities at imaginary frequencies
and the Li-Li ($2s+2s$) and  Rb--Rb ($5s+5s$) $C_6$ and $C_8$ coefficients were determined earlier
(see, e.g.,~\cite{PorDer03,DerPorBab10} and references therein).
The ground state Yb--Yb $C_6$ and $C_8$ coefficients as well as the dynamic electric dipole and electric quadrupole
polarizabilities of the $^1\!S_0$ state at imaginary frequencies were obtained in our recent work~\cite{PorSafDer13l}.
These values are compiled in Table~\ref{C8aB} for reference and comparison with selected other results
\cite{PucKcePac11,PucKcePac12e,SafSaf11,KitEnoKas08l,BruHut13}. We use the dynamic
polarizabilities from these works to determine the van der Waals coefficients for Yb--Li ($^1\!S_0 + 2s$) and
Yb--Rb ($^1\!S_0 + 5s$) dimers.

The $C_8(1,2)$ and $C_8(2,1)$ values as well as the final value of the $C_8$ coefficient
for the Yb--Rb ($^1\!S_0 + 5s$) dimer are given in Table~\ref{C8aB}. The final value
of the $C_8$ coefficient for ($^1\!S_0 + 2s$) Yb--Li dimer is also listed in Table~\ref{C8aB}. In an alternative approach,
we also carried out the calculation using the approximate formula~\eref{C8approx}
and obtained $C_8 \approx 1.27 \times 10^5$~a.u. and $C_8 \approx 3.20 \times 10^5$~a.u., for Yb--Li and Yb--Rb, respectively.
 These values are identical to our values obtained with
 Eqs.~(\ref{vdW}) and (\ref{C_12}). Our $C_6$ value is in agreement with a very recent accurate analysis of the
 photassociation data \cite{BorZucCiu13}, while our $C_8$ result is lower by 3$\sigma$ than  Ref.~\cite{BorZucCiu13} value.

We substitute the static polarizability uncertainties listed in~\tref{C8aB} into
Eqs.~(\ref{C_AB_l}) and (\ref{C_AB}) to estimate the uncertainty of the $C_8(^1\!S_0+5s)$ coefficient.
The evaluation of the polarizability uncertainties was discussed in detail in Ref.~\cite{PorSafDer13l}.
The resulting uncertainties in the Yb-Li $C_8(^1\!S_0+2s)$ and Yb-Rb $C_8(^1\!S_0+5s)$ coefficients is estimated to be 2\%.
\subsection{Yb--Rb ($6s^2\, ^1\!S_0 + 5p_{1/2}$) dimer}
\label{5p1S0}
The $C_6$ coefficient for the $6s^2\, ^1\!S_0 + 5p_{1/2}$ dimer cannot be calculated using
Eqs.~(\ref{vdW}) and~(\ref{CAB_11}) due to the downward $5p_{1/2} - 5s$ transition in Rb. We derive the
expression for this $C_6$ coefficient below.
In this subsection, we designate $A \equiv\, 5p_{1/2}$ and $B \equiv\, ^1\!S_0$.

We start from the general formula,~\eref{C6_YbRb}, which in this case is reduced to
\begin{eqnarray}
C^{AB}_{6}(\Omega ) = \sum_{j=1/2}^{3/2} A_{j 1}(\Omega) X_{j 1} .
\label{Eq:5p1S0}
\end{eqnarray}

Since the projection of the Rb total angular momentum $M_A=1/2$ and
the projection of the Yb total angular momentum $M_B=0$, the only possible value of
$\Omega = M_A+M_B = 1/2$.
Substituting $J_A=M_A=1/2$, $J_B=M_B=0$, and $J=1$ into~\eref{AjJ}
and setting $\Omega = 1/2$, we obtain $A_{j 1}(\Omega=1/2) = 1/3$ for both possible
values $j=1/2$ and $j=3/2$.

Furthermore, all states with $j=3/2$ are above the $5p_{1/2}$ state, i.e.,
there are no downward transitions from the $5p_{1/2}$ state to any state with  $j=3/2$.
Then, using Eqs.~(\ref{XjJ}) and (\ref{CP}), we obtain for $X_{\frac{3}{2} 1}$:
\begin{eqnarray}
X_{\frac{3}{2} 1} = \frac{9}{\pi} \int_0^\infty \alpha^A_{1 \frac{3}{2}}(i\omega)\,
\alpha^B_1(i\omega )\, d\omega ,
\label{X31}
\end{eqnarray}%
where  $\alpha^A_{1 j}(i\omega)$ is the part of the dynamic electric dipole
$5p_{1/2}$ polarizability at the imaginary frequency with $J_n=j$:
\begin{equation}
\alpha^A_{1 j}(i\omega) = \frac{1}{3} \sum_{\gamma_n} \frac{(E_n -E_A) |\langle \gamma_n J_n=j ||d|| A \rangle|^2}
{(E_n - E_A)^2 + \omega^2}
\label{alpha_a1}
\end{equation}
and $\alpha_1^B$ is the dynamic electric-dipole polarizability of the $^1\!S_0$ state:
\begin{equation}
\alpha^B_1(i\omega) = \frac{2}{3} \sum_k \frac{(E_k - E_B) |\langle
k ||d|| ^1\!S_0 \rangle|^2} {(E_k - E_B)^2 + \omega^2} .
\label{alpha_B1}
\end{equation}

The case of $j=1/2$ is more complicated because there is the
downward $5p_{1/2} - 5s$ transition, precluding direct application of the Casimir-Polder identity (\ref{CP}).
Using Eq.~(\ref{XjJ}) we can separate out the contribution of the $5s$ state from the sum over $n$. Then we obtain
\begin{eqnarray}
X_{\frac{1}{2} 1} &\equiv& X^{(1)}_{\frac{1}{2} 1} + X^{(2)}_{\frac{1}{2} 1} =   \\
&=& |\langle A ||d|| 5s \rangle|^2 \, \sum_k \frac{|\langle B ||d|| k
\rangle|^2} {-\omega_{As} + \omega_{kB}} \nonumber \\
&+& \sum_{\gamma_n \neq 5\!s, k} \frac{|\langle A ||d|| \gamma_n J_n\!=\!1/2 \rangle|^2 \,
|\langle B ||d|| k \rangle|^2} {\omega_{nA} + \omega_{kB}} \nonumber,
\end{eqnarray}
where $\omega_{As} \equiv E_A-E_{5s}$, $\omega_{nA} \equiv E_n-E_A$
and $\omega_{kB} \equiv E_k-E_B$. Both frequencies $\omega_{nA}$ and
$\omega_{kB}$ in the second term, $X^{(2)}_{\frac{1}{2} 1}$, are positive for
any $n$ and $k$. Using Eq.~(\ref{CP}), we can represent this term by
\begin{eqnarray}
X^{(2)}_{\frac{1}{2} 1} &=& \frac{2}{\pi} \int_0^\infty d\omega \sum_{\gamma_n \neq 5s}
\frac{\omega_{nA} \, |\langle A ||d|| \gamma_n J_n=1/2 \rangle|^2}
{\omega_{nA}^2+\omega^2} \nonumber \\
&\times& \sum_{k} \frac{\omega_{kB} \, |\langle B ||d|| k
\rangle|^2} {\omega_{kB}^2+\omega^2} .
\label{X2_11}
\end{eqnarray}
Next, we add and subtract the term $|\gamma_n\rangle = |5s\rangle$ into the sum over
$\gamma_n$ under the integral in Eq.~(\ref{X2_11}) and use again Eq.~(\ref{CP}):
\begin{eqnarray}
X^{(2)}_{\frac{1}{2} 1} &=& \frac{9}{\pi} \int_0^\infty \alpha^A_{1 \frac{1}{2}}(i\omega)\,
\alpha^B_1(i\omega )\, d\omega \nonumber \\
&+& |\langle A ||d|| 5s \rangle|^2
\sum_{k} \frac{|\langle B ||d|| k \rangle|^2} {\omega_{kB} + \omega_{As}} .
\label{X211}
\end{eqnarray}%

Combining $X^{(1)}_{\frac{1}{2} 1}$ and $X^{(2)}_{\frac{1}{2} 1}$, we arrive at
\begin{eqnarray}
X_{\frac{1}{2} 1} &=& \frac{9}{\pi} \int_0^\infty \alpha^A_{1 \frac{1}{2}}(i\omega)\,
\alpha^B_1(i\omega )\, d\omega \nonumber \\
&+& 3\, |\langle A ||d|| 5s \rangle|^2 \alpha^B_1(\omega_{As}),
\label{X11}
\end{eqnarray}%
where
\begin{equation}
\alpha^B_1(\omega_{As}) = \frac{2}{3} \sum_k \frac{\omega_{kB}
|\langle k ||d|| B \rangle|^2} {\omega_{kB}^2 - \omega_{As}^2} .
\label{alpha_B19}
\end{equation}

Taking into account that $A_{\frac{1}{2} 1} = A_{\frac{3}{2} 1} = 1/3$,
and substituting the expressions for $X_{j1}$ into  Eq.~(\ref{Eq:5p1S0}), we arrive at the final expression
for $C^{AB}_6$ in the present case:
\begin{eqnarray}
C^{AB}_6 &=& \frac{3}{\pi} \int_0^\infty \alpha^A_1 (i\omega)\, \alpha^B_1(i\omega)\,
d\omega \nonumber \\
&+& |\langle 5p_{1/2} ||d|| 5s \rangle|^2 \, \alpha^B_1(\omega_{As}) ,
\label{C6_5p1S0}
\end{eqnarray}
where $\alpha^A_1(i\omega)$ is the dynamic electric dipole polarizability of
the $5p_{1/2}$ state at the imaginary argument $i \omega$.

The quantities needed to calculate the $C_6(^1\!S_0 + 5p_{1/2})$ coefficient and to evaluate
its uncertainty are summarized in~\tref{Tab:1S0_5p}.
\begin{table}[tbp]
\caption{The quantities (in a.u.) used to calculate the $C_6(^1\!S_0 + 5p_{1/2})$ coefficient and
to estimate its uncertainty. $\alpha_1(0)$ is the static polarizability,  $\alpha_1(\omega_{As})$ is the
Yb polarizability calculated at the real frequency $\omega_{As} \equiv E_{5p_{1/2}} - E_{5s}$.
The uncertainties are given in parenthesis.}
\label{Tab:1S0_5p}%
\begin{ruledtabular}
\begin{tabular}{lccll}
       &   State            &   Quantity                          &  Results                  &     \\
\hline
Rb     &   $5p_{1/2}$       & $\alpha_1(0)$                       & 810.8(8)\footnotemark[1]  &  Theory + Exp. \\
       &                    & $|\langle 5p_{1/2}||d||5s \rangle|$ & 4.228(6)\footnotemark[2]  &  Experiment\\[0.3pc]

Yb     &   $^1\!S_0$        & $\alpha_1(0)$                       & 141(2)\footnotemark[3]    &  Theory \\
       &   $^1\!S_0$        & $\alpha_1(\omega_{As})$             & 183(3)                    &  This work \\[0.3pc]

Yb--Rb & $^1\!S_0+5p_{1/2}$ & $C_6$                               & 7607(114)                 &  This work \\
       & $^1\!S_0+5p_{1/2}$ & $C_6$                               & 5684(98)\footnotemark[4]  &  Experiment
\end{tabular}
\end{ruledtabular}
\footnotemark[1]{Ref.~\cite{ZhuDalPor04}};
\footnotemark[2]{Ref.~\cite{GutAmiFio02l}};
\footnotemark[3]{Ref.~\cite{SafPorCla12}};
\footnotemark[4]{Ref.~\cite{NemBauMun09l}}.
\end{table}
The Rb dynamic $5p_{1/2}$ polarizability was calculated in~\cite{ZhuDalPor04} in the framework of
DF + MBPT approximation~\cite{DzuFlaKoz96b}.
The Yb dynamic electric-dipole $^1\!S_0$ polarizabilities at the imaginary frequencies were calculated in~\cite{SafPorCla12}.
Using experimental and theoretical data, the static electric-dipole polarizability of the $5p_{1/2}$ state was
found to be 810.8(8) a.u.~\cite{ZhuDalPor04}. Using these polarizabilities
and evaluating the integral using the finite sum~\eref{finsum}, we obtain from~\eref{C6_5p1S0}
\begin{eqnarray}
\frac{3}{\pi} \int_0^\infty \alpha^A_1 (i\omega)\, \alpha^B_1(i\omega )\, d\omega &\approx& 4336~\textrm{a.u.}
\end{eqnarray}%

We use the experimental value for the matrix element
$|\langle 5p_{1/2} ||d|| 5s \rangle| = 4.228(6)$ a.u.~\cite{GutAmiFio02l} in the the second term of~\eref{C6_5p1S0}.
The quantity $\alpha^B_1(\omega_{As})$ is the  electric-dipole ground-state Yb $^1\!S_0$ polarizability at the real frequency
$\omega_{As} \equiv E_{5p_{1/2}} - E_{5s}$. Adding core and valence contributions (see \cite{SafPorCla12} for detail), we obtain
\begin{eqnarray}
\alpha^B_1(\omega_{As}) &=& \alpha^{B({\rm val})}_1 + \alpha^{B(\mathrm{core)}}_1 \nonumber \\
&\approx& 177 + 6 = 183 \,\, \mathrm{a.u.} .
\label{om_As}
\end{eqnarray}%
Thus, the contribution of the second term in~\eref{C6_5p1S0} to the $C_6$ coefficient is
$$|\langle 5p_{1/2} ||d|| 5s \rangle|^2 \, \alpha^B_1(\omega_{As}) \approx 3271~\textrm{a.u.}$$
We note that both terms are comparable in their magnitude. Adding these terms,
we obtain $C_6 \approx 7607~\textrm{a.u.}$.

The uncertainty of this $C_6$ coefficient can be evaluated using Eqs.~(\ref{C_AB_l}) and (\ref{C_AB}).
Taking into account that the fractional uncertainty of the Rb static $5p_{1/2}$
polarizability, 0.1\%, is negligible in comparison to the uncertainty of the Yb static $^1\!S_0$ polarizability, 1.5\%,
the accuracy of the first term in~\eref{C6_5p1S0} (4336 a.u.) is dominated by the
the accuracy of the $^1\!S_0$ polarizability, i.e. 1.5\%.
In the second term (3271 a.u.), the fractional uncertainty of the matrix element
$\langle 5p_{1/2} ||d|| 5s \rangle$, 0.14\%, is negligible in comparison with the
fractional uncertainty of the dynamic polarizability
$\delta \alpha^B_1(\omega_{sp})\approx \delta \alpha^B_1(0) \approx$ 1.5\%.
Thus, we assume that the uncertainty in the second term is 1.5\% and the
final uncertainty for the $C_6$ coefficient is also 1.5\%;
$C_6 = 7607(114)$.
The $C_6(^1S_0+5p_{1/2})$ coefficient is 33\% larger than the value obtained from the fit of the photoassociation data
with Leroy-Berstein method \cite{NemBauMun09l}. However, our $C_6(^1S_0+5s)$ ground-state value of 2837(57)~a.u. is 14\% larger
than result of a similar fit that yielded 2485(21) a.u. \cite{MunBruMad11}. Very recent accurate analysis of the photassociation
data gives 2837(13)~a.u.~\cite{BorZucCiu13}, which is in perfect agreement with our central value. This may indicate that the
uncertainties of the values obtained by the experimental data fit with commonly used Leroy-Bernstein method may be larger than
expected, especially for the exited states.

\subsection{Yb--Rb ($6s6p\, ^3\!P_1^o + 5s$) dimer}
\label{5s3P1}
This case also requires special attention because
there is the downward $^3\!P_1^o -\, ^1\!S_0$
transition in Yb. In this subsection, we designate
$A \equiv\, 5s$ and $B \equiv\, 6s6p\,^3\!P_1^o$.

The general expression for the $C_{6}$ coefficient, given by~\eref{C6_YbRb},
leads to
\[
C_{6}(\Omega )=\sum_{j=1/2}^{3/2} \sum_{J=0}^{2} A_{j J}(\Omega) X_{j J} .
\]
Taking into account that $M_A=1/2$ and $M_B=0,1$, the possible
values of $\Omega$ are 1/2 and 3/2, where $M_B=0$ corresponds to $\Omega=1/2$
and $M_B=1$ corresponds to $\Omega=3/2$.

The coefficients $A_{j J}(\Omega)$ are given by~\eref{AjJ}.
Their calculation is straightforward and numerical values are listed in Table~\ref{A_J}
for different values of $j$, $J$, and $\Omega$.
\begin{table}[tbp]
\caption{The values of the $A_{j J}(\Omega)$ and $X_{j J}$ coefficients for
different $j$, $J$, and $\Omega$. The contributions to the $C_6$ coefficients are listed
in the columns $C_6(\Omega=1/2)$ and $C_6(\Omega=3/2)$.
 The contributions of
$\delta X_{1/2,0}$ and $\delta X_{3/2,0}$ are given in the rows $\delta_{1/2}$ and $\delta_{3/2}$; they are included in
the terms $X_{1/2,0}$ and $X_{3/2,0}$, respectively.  Final values of the $C_6$ coefficients and their uncertainties
(in parentheses) are presented in the row labeled ``Total''.}
\label{A_J}%
\begin{ruledtabular}
\begin{tabular}{clccccc}
$j$  & $J$ & \multicolumn{2}{c}{$A_{j J}$}&\multicolumn{1}{c}{$X_{j J}$} & \multicolumn{2}{c}{$C_6$} \\
     &     & $\Omega=1/2$ & $\Omega=3/2$  &                              & $\Omega=1/2$ & $\Omega=3/2$ \\
\hline
$\delta_{1/2}$
     &  0  &              &               &       $-$189                 &              &          \\
1/2  &  0  &    2/9       &      0        &          680                 &   151        &   0      \\
1/2  &  1  &    1/18      &     1/9       &         4887                 &   272        &  543     \\
1/2  &  2  &   11/90      &     2/15      &         7419                 &   907        &  989      \\[0.3pc]

$\delta_{3/2}$
     &  0  &              &               &       $-$399                 &              &          \\
3/2  &  0  &    2/9       &     1/12      &         1327                 &   295        &  111     \\
3/2  &  1  &    1/18      &    11/72      &         9671                 &   537        & 1478     \\
3/2  &  2  &   11/90      &    11/120     &        14675                 &  1794        & 1345     \\
Total&     &              &               &                              &  3955(160)        & 4466(180)     \\
\end{tabular}
\end{ruledtabular}
\end{table}

Starting from the general expression for $X_{j J}$ given by~\eref{XjJ} and using the approach
discussed in the previous subsection, we obtain
\begin{eqnarray}
X_{j J} = \frac{27}{\pi} \int_0^\infty \alpha^A_{1 j}(i\omega)\,
\alpha^B_{1 J}(i\omega )\, d\omega + \delta X_{j J}\,\delta _{J 0} ,
\label{XJ}
\end{eqnarray}%
where $\alpha_{1 j}^A(i \omega)$ is a contribution to the Rb $5s$ electric dipole
polarizability given by~\eref{alpha_a1} (with $A=5s$) and $\alpha_{1 J}^B (i \omega)$ is a
contribution to the scalar part of the Yb electric-dipole $^3\!P_1^o$ polarizability,
determined as
\begin{equation}
\alpha^B_{1 J}(i\omega) = \frac{2}{9} \sum_{\gamma_k} \frac{(E_k - E_B)
|\langle \gamma_k J_k=J ||d|| B \rangle|^2} {(E_k - E_B)^2 + \omega^2} .
\label{alpha_B2}
\end{equation}

The correction $\delta X_{j 0}$ to the $X_{j 0}$ term is due to the downward
$^3\!P_1^o \rightarrow \,^1\!S_0$ transition. One can show that it can be written as
\begin{eqnarray}
\delta X_{j 0} &=& 2\,|\langle ^{3}\!P_{1}^{o}||d||^{1}\!S_{0}\rangle|^{2} \nonumber \\
&\times& \sum_{\gamma_n} \frac{(E_{n}-E_A)\, |\langle \gamma_n,J_n=j||d||A\rangle|^2}
{(E_{n}-E_A)^{2}-\omega _{0}^{2}} ,
\label{delX0}
\end{eqnarray}%
where $\omega_0 \equiv E_{^{3}\!P_{1}^{o}}-E_{^{1}\!S_{0}}$ and the total
angular momentum of the intermediate states $J_n$ is fixed and equal to $j$. In our case, $j=1/2$
or 3/2.

The Rb dynamic $5s$ polarizabilities at imaginary frequencies were obtained in Ref.~\cite{DerPorBab10}.
We calculated  $\delta X_{j 0}$ following the approach discussed in~\cite{PorDer03,ZhuDalPor04}.
The calculation of the scalar part of the  dynamic electric dipole $^3\!P_1^o$ polarizability
was discussed in detail in~\cite{PorSafDer13l}.
The resulting contributions to $C_6$ coefficient are listed in Table~\ref{A_J}.

We note that $\delta X_{j 0} < 0$ for both $j=1/2$ and $j=3/2$. As follows from~\eref{delX0},
we need to calculate the dynamic $5s$ polarizability at the frequency
$\omega_0 = E_{^3\!P_1^o}-E_{^1\!S_0} \approx 0.082$ a.u. to determine $\delta X_{j 0}$. The dominant contribution
to this polarizability comes from the $5p_{1/2, 3/2}$ states. Since
$E_{5p_j}-E_{5s} \approx 0.057$ a.u., the energy denominators ($E_{5p_j}-E_{5s} - \omega_0$)
will be small and negative. It leads to the negative values of $\alpha_{1j}^{5s}(\omega_0)$ and,
subsequently, $\delta X_{j 0}$, for both $j=1/2$ and $j=3/2$.

Since the uncertainty of the Rb static $5s$ polarizability, 0.2\%, is negligible in comparison with the uncertainty of the
Yb static scalar $^3\!P_1^o$ polarizability (3.5\%), the latter determines the uncertainty
of the $C_6$ coefficient for the $(^3\!P_1^o + 5s)$ dimer. Our final values are $C_6(\Omega=1/2)=3955(160)$
and $C_6(\Omega=3/2)=4466(180)$. Clearly at long range the $\Omega=3/2$ potential is more attractive than the  $\Omega=1/2$  potential.

\section{Conclusion}
To summarize, in this work we obtained accurate $C_6$ and $C_8$ values for the Yb-Rb and Yb-Li dimers
of particular experimental interest which are needed for efficient production, cooling, and control of molecules.
For the case when $A$ and $B$ are  spherically symmetric atomic states
and there are no downward transitions from either of these states,
we derived a semi-empirical formula for the $C_8$ coefficient of
heteronuclear ($A+B$) dimers. We evaluated the $C_8$ coefficient for the Yb--Li $^1\!S_0+2s$ and  Yb--Rb $^1\!S_0+5s$ dimers
using the exact and approximate expressions and found excellent agreement between these values.
Our calculations of $C_8$ coefficients will allow accurate extraction of $C_6$ from the photoassociation spectra and may allow
to estimate contribution of the $C_{10}$ in the interaction potential. We performed detailed uncertainty analysis and provided
stringent bounds on all of the quantities calculated in this work to allow future benchmark tests of experimental methodologies
and theoretical molecular models.

\section*{Acknowledgement}
We thank P. Julienne and T. Porto for helpful discussions.
This research was performed under the sponsorship of the
U.S. Department of Commerce, National Institute of Standards and
Technology, and was supported by the National Science Foundation
under Physics Frontiers Center Grant No. PHY-0822671 and by the
Office of Naval Research. The work of S.G.P. was supported in part by
US NSF Grant No.\ PHY-1212442.
The work of A.D. was supported in part by the US NSF Grant No. PHY-1212482.

\end{document}